\documentclass[aps,twocolumn]{revtex4}

\begin{document}

\title
{Non-Hamiltonian Commutators in Quantum Mechanics}

\author{Alessandro Sergi
\footnote{ (E-mail: asergi@unime.it)}}
\affiliation{
Dipartimento di Fisica, Sezione Fisica Teorica,
Universit\'a degli Studi di Messina,
Contrada Papardo C.P. 50-98166 Messina, Italy
}

\begin{abstract}
The symplectic structure of quantum commutators is first unveiled and then exploited
to introduce generalized non-Hamiltonian brackets in quantum mechanics.
It is easily recognized that quantum-classical systems
are described by a particular realization of such a bracket.
In light of previous work, this introduces
a unified approach to classical and
quantum-classical non-Hamiltonian dynamics.
In order to illustrate the use
of non-Hamiltonian commutators, it is shown how to define
thermodynamic constraints in quantum-classical systems.
In particular, quantum-classical Nos\'e-Hoover equations of motion 
and the associated stationary density matrix are derived.
The non-Hamiltonian commutators for both Nos\'e-Hoover chains
and Nos\'e-Andersen (constant-pressure constant temperature)
dynamics are also given.
Perspectives of the formalism are discussed.
\end{abstract}

\maketitle

{\bf Submitted to Phys. Rev. E on August 8 2005}

\section{Introduction}

In order to describe phenomena in the real world,
classical and quantum systems are represented by means of 
Hamiltonian mathematical theories~\cite{goldstein,mccauley,dirac,balescu}.
However, when studying systems with many degrees of freedom,
the need of performing numerical calculations on computers 
has led to the development of non-Hamiltonian
mathematical structures~\cite{andersen,nose,hoover}.

In the classical case, non-Hamiltonian formalisms are
typically employed to implement 
thermodynamic constraints~\cite{ferrario}
using just few additional degrees of freedom
(whereas by using Hamiltonian theories one should resort to
an infinite number of degrees of freedom).
Just recently, it has been shown that
the non-Hamiltonian dynamics of classical systems 
can be formulated in a unified way by means of
generalized brackets which ensure energy conservation~\cite{sergi,sergi2}
and subsumes Dirac's formalism~\cite{dirac} for
systems with holonomic constraints~\cite{sergi3}.
Other approaches to classical non-Hamiltonian brackets
can be found in Refs.~\cite{edwards,tarasov}.

In the quantum case, the impossibility to solve on computers
full quantum dynamics for interacting many-body systems
has led to the development of quantum-classical theories.
Indeed, a generalized bracket to treat quantum-classical systems
has been  proposed by various authors~\cite{qc-bracket}.

Since in the classical case non-Hamiltonian brackets
are obtained by modifying the symplectic structure
of the Poisson bracket~\cite{sergi,sergi2}, 
in order to deal with the quantum case
one could first make apparent the symplectic structure
of the commutator (which is the Hamiltonian bracket of quantum mechanics)
and then generalize it in order to obtain
a non-Hamiltonian quantum bracket (commutator).
In this paper it is shown that this is indeed possible.
The non-Hamiltonian commutator, which is obtained by this procedure,
is then used to reformulate quantum-classical
brackets~\cite{qc-bracket}.
Thus, it is stressed that quantum-classical dynamics
can be regarded as a form of non-Hamiltonian
quantum mechanics because the quantum-classical
bracket does not satisfy the Jacobi relation and,
as a consequence, the time-translation
invariance of the algebra is violated.
In order to illustrate the use
of non-Hamiltonian commutators, it is shown how to define
thermodynamic constraints in quantum-classical systems.
The particular case of the Nos\'e-Hover thermostat~\cite{nose,hoover} 
is treated in full details
and the associated stationary density matrix is derived.
The more general cases of Nos\'e-Hover chains~\cite{nhc}
and constant pressure and temperature~\cite{andersen,nose,ferrario}
bring no major difference neither conceptually nor
technically and are treated in less detail. 
It is worth noting that some past attempts of introducing
Nos\'e-Hover dynamics in quantum calculations~\cite{tosatti,schnack}
used a simpler form of quantum-classical dynamics
which did not treat correctly the quantum back-reaction 
on the classical variables.
The possibility of applying thermodynamic constraints
to quantum-classical dynamics is a technical advance
that could lead to further theoretical and computational
achievements with regards to the study
of open quantum systems~\cite{davis,weiss}.
With respect to this, a non-trivial major obstacle
is the development of efficient algorithms
to simulate long-time quantum dynamics.

Besides the technical applications of non-Hamiltonian
commutators to the particular case of quantum-classical 
dynamics, one could appreciate on a more conceptual level
that, in light of previous work, non-Hamiltonian brackets
provide a unified approach to non-Hamiltonian dynamics both
in the classical and quantum case.
In addition, if one is willing to indulge in speculations,
it is worth to note that the mathematical structure presented in this paper
may be shown to generalize the formalisms that
a number of authors have already presented 
in the literature~\cite{qm-structure,nambu,weinberg,jones,birula}.
In particular it is worth mentioning that non-Hamiltonian
commutators could be used, in principle, in order to introduce
non-linear effects in quantum mechanics along the lines already proposed
by Weinberg~\cite{weinberg}.
Therefore, one could foresee interesting applications of non-Hamiltonian
commutators in various fields.

The paper is organized as follows:
in section~\ref{sec:matrix} the symplectic structure of 
Hamiltonian quantum mechanics is unveiled and 
its generalization by means of the non-Hamiltonian commutator
is proposed.
In section~\ref{sec:bracket} it is shown that 
the quantum-classical bracket can be written
in matrix form as a non-Hamiltonian commutator.
Such a form easily illustrates
the failure of the Jacobi relation~\cite{kapral2}.
In section~\ref{sec:nose} non-Hamiltonian commutators 
for quantum-classical systems are used in order to introduce,
following Refs.~\cite{sergi,sergi2}, Nos\'e thermostatted dynamics
on the classical degrees of freedom.
Nos\'e-Hoover chains and constant pressure constant
temperature (NPT) equations of motion bring no major
difference and are treated in less detail
in appendix~\ref{sec:nhc-npt}.
In section~\ref{sec:stationary} it is proven
that the stationary density matrix 
under the quantum-classical Nos\'e-Hoover
equation of motion exists and 
its explicit form is given up to order $\hbar$.
Conclusions and perspectives are given in the final section.

\section{Non-Hamiltonian Quantum Mechanics} \label{sec:matrix}

It is well-known that classical and quantum dynamics
share an analogous algebraic structure~\cite{dirac,balescu} realized
by means of specific brackets: Poisson brackets in the classical case
and commutators in the quantum one.
It is also known that Poisson brackets have a symplectic
structure that is easily represented in matrix 
form~\cite{goldstein,mccauley}.
If one denotes the point in phase space as $X=(R,P)$,
where $R$ and $P$ are the usual coordinates and momenta respectively,
by defining the antisymmetric matrix
\begin{equation}
\mbox{\boldmath$\cal B$}=\left[\begin{array}{cc}0 & 1\\ -1 & 0\end{array}\right]
\label{B}
\end{equation}
the Poisson bracket of any two phase space function $a(X)$ 
and $b(X)$ can be written in matrix form as
\begin{equation}
\{a,b\}=\sum_{i,j=1}^{2N}\frac{\partial a}{\partial X_i}{\cal B}_{i j}\frac{\partial 
b}{\partial X_j},
\label{eq:symplectic-bracket}
\end{equation}
where $2N$ is phase space dimension.
In Refs.~\cite{sergi,sergi2,sergi3}, Eq.~(\ref{eq:symplectic-bracket}) 
has been generalized introducing an antisymmetric tensor field
${\cal B}_{i j}(X)=-{\cal B}_{j i}(X)$ so that a general bracket
$\{\ldots,\ldots\}_X$,
having the same matrix structure of that in 
Eq.~(\ref{eq:symplectic-bracket}),
could be introduced and non-Hamiltonian equations of motion
could be written as
\begin{equation}
\dot{X}_i=\{X_i,{\cal H}\}_X=\sum_{j=1}^{2N}
{\cal B}_{i j}(X)\frac{\partial{\cal H}}{\partial X_j}\;,
\end{equation}
where $\cal H$ is the ``Hamiltonian'' or generalized energy.

As one could expect, the commutator can also be written in matrix form
using the symplectic structure of Eq.~(\ref{B}).
If one considers a set of quantum variables $\hat{\chi}_{\alpha}$, $\alpha=1,...,n$,
which can be canonical, non-canonical or anti-commuting variables,
the commutator 
$[\hat{\chi}_{\alpha},\hat{\chi}_{\nu}]=\hat{\chi}_{\alpha}\hat{\chi}_{\nu}
-\hat{\chi}_{\nu}\hat{\chi}_{\alpha}$ ($\alpha,\nu=1,...,n$)
can be expressed as
\begin{equation}
[\hat{\chi}_{\alpha},\hat{\chi}_{\nu}]=
\left[\begin{array}{cc}\hat{\chi}_{\alpha} & \hat{\chi}_{\nu}\end{array}\right]
\cdot\left[\begin{array}{cc}0 & 1\\ -1 & 0\end{array}\right]\cdot
\left[\begin{array}{c}\hat{\chi}_{\alpha} \\ \hat{\chi}_{\nu}\end{array}\right]\;.
\label{eq:quantum-algebra}
\end{equation}
The above matrix form of the commutator permits to appreciate the common
symplectic structure of both classical and quantum mechanics.

Given the Hamiltonian operator $\hat{H}$ of the system,
the law of motion in the Heisenberg picture can also be written in matrix 
form as
\begin{equation}
 \frac{d\hat{\chi}_{\alpha}}{dt}=\frac{i}{\hbar}
\left[\begin{array}{cc} \hat{H} & \hat{\chi}_{\alpha} \end{array}\right]
\cdot\mbox{\boldmath$\cal B$}\cdot
\left[\begin{array}{c} \hat{H} \\ \hat{\chi}_{\alpha}\end{array}\right]
=i\hat{\cal L}\hat{\chi}_{\alpha}\;,
\label{qlm}
\end{equation}
where it has been introduced the Liouville operator
\begin{equation}
i \hat{\cal L}=\frac{i}{\hbar}
\left[\begin{array}{cc} \hat{H} & \ldots \end{array}\right]
\cdot\mbox{\boldmath$\cal B$}\cdot
\left[\begin{array}{c} \hat{H} \\ \ldots \end{array}\right]\;.
\end{equation}

The algebra of commutators is a Lie algebra.
This means in particular that the commutator satisfies the following properties:
\begin{eqnarray}
[ \hat{\chi}_{\alpha} , \hat{\chi}_{\nu} ]&=&
-[ \hat{\chi}_{\nu} , \hat{\chi}_{\alpha} ] \label{antisymmetry} \\
\left[\hat{\chi}_{\alpha} \hat{\chi}_{\nu},\hat{\chi}_{\sigma}\right]&=&
\hat{\chi}_{\alpha}[\hat{\chi}_{\nu},\hat{\chi}_{\sigma}]+
[\hat{\chi}_{\alpha},\hat{\chi}_{\sigma}]\hat{\chi}_{\nu}
\\
\left[c,\hat{\chi}_{\nu}\right]&=& 0
\label{lie-last},
\end{eqnarray}
where $c$ is a so called c-number and $\alpha,\nu,\sigma=1,...,n$.
Besides properties in Eqs.~(\ref{antisymmetry}-\ref{lie-last}),
in order to have a Lie algebra, it is necessary that the so called Jacobi 
identity holds
\begin{equation}
{\cal J}=[\hat{\chi}_{\alpha},[\hat{\chi}_{\nu},\hat{\chi}_{\sigma}]]+
[\hat{\chi}_{\sigma},[\hat{\chi}_{\alpha},\hat{\chi}_{\nu}]]+
[\hat{\chi}_{\nu},[\hat{\chi}_{\sigma},\hat{\chi}_{\alpha}]]
=0 \label{jacobi}.
\end{equation}
The Jacobi identity ensures that the algebra is invariant under the
law of motion and as such it states an integrability condition.
In the above formalism it can be appreciated that
the antisymmetry of the commutator~(\ref{antisymmetry}) arises from
the antisymmetry of the symplectic matrix $\mbox{\boldmath$\cal B$}$ 
and ensures that if $\hat{H}$ is not 
explicitly time-dependent then it is a constant of motion
\begin{equation}
\frac{d}{dt}\hat{H}= i \hat{\cal L}\hat{H}=0\;.
\end{equation}
The conservation of energy under time-translation
defined by means of antisymmetric brackets
is another nice property shared both by the algebra of Poisson brackets
on classical phase space and by the algebra of commutators of quantum variables.

Using the operator language of Eq.~(\ref{eq:quantum-algebra}),
one can define a generalized commutator as
\begin{equation}
[\hat{\chi}_{\alpha},\hat{\chi}_{\nu}]=
\left[\begin{array}{cc}\hat{\chi}_{\alpha} & \hat{\chi}_{\nu}\end{array}\right]
\cdot\mbox{\boldmath$\cal D$}\cdot
\left[\begin{array}{c}\hat{\chi}_{\alpha} \\ \hat{\chi}_{\nu}\end{array}\right]\;,
\label{eq:gen-quantum-algebra}
\end{equation}
where $\mbox{\boldmath$\cal D$}$ is an antisymmetric matrix operator
of the form
\begin{equation}
\mbox{\boldmath$\cal D$}=
\left[\begin{array}{cc}0 & \hat{\zeta}\\ -\hat{\zeta} & 
0\end{array}\right]\;,
\end{equation}
with $\hat{\zeta}$ arbitrary operator or c-number.
Generalized equations of motion could then be defined as
\begin{equation}
 \frac{d\hat{\chi}_{\alpha}}{dt}=\frac{i}{\hbar}
\left[\begin{array}{cc} \hat{H} & \hat{\chi}_{\alpha} \end{array}\right]
\cdot\mbox{\boldmath$\cal D$}\cdot
\left[\begin{array}{c} \hat{H} \\ \hat{\chi}_{\alpha}\end{array}\right]
=i\hat{\cal L}\hat{\chi}_{\alpha}.
\label{eq:gen-qlm}
\end{equation}
It must be stressed that the non-Hamiltonian commutator defined in 
Eq.~(\ref{eq:gen-quantum-algebra})
could violate the Jacobi relation~(\ref{jacobi})
so that in general it does not define a Lie algebra.
The non-Hamiltonian commutator of Eq.~(\ref{eq:gen-quantum-algebra})
defines, of course, a generalized form of quantum mechanics.
However, in this generalized theory the Hamiltonian operator $\hat{H}$
is still a constant of motion because of the antisymmetry of 
$\mbox{\boldmath$\cal D$}$.
It is interesting to note that $\mbox{\boldmath$\cal D$}$ could in principle
depend from the quantum variables $\hat{\chi}_{\alpha}$.
Then Eq.~(\ref{eq:gen-qlm}) can be thought of as a generalization
to the Heisenberg picture of
the mathematical formalism proposed by Weinberg~\cite{weinberg}
in order to introduce non-linear effects in quantum mechanics.

In the next section it will be shown that
the non-Hamiltonian commutator defined in Eq.~(\ref{eq:gen-quantum-algebra})
and the non-Hamiltonian equations of motion~(\ref{eq:gen-qlm})
provide the mathematical structure for 
quantum-classical evolution~\cite{qc-bracket}.

\section{Non-Hamiltonian Commutators in
 Quantum-Classical Mechanics}\label{sec:bracket}

Quantum-classical systems can be treated by means of an
algebraic approach.  This has been already proposed by 
a number of authors~\cite{qc-bracket}
by means of a quantum-classical bracket which
does not satisfy the Jacobi relation.
A quantum-classical system is composed of
both quantum $\hat{\chi}$ and classical $X$ degrees
of freedom. 
The quantum variables depends from the classical point $X$
so that an abstract space is defined in such a way
that a Hilbert space (where quantum dynamics takes place)
is attached to each phase space point.
In turn, a displacement of the phase space point
determines a consistent effect on quantum evolution
in the Hilbert space.
The energy of the system is defined in terms of
a quantum-classical Hamiltonian operator $\hat{H}=\hat{H}(X)$ 
coupling quantum and classical variables
$E={\rm Tr}'\int dX \hat{H}(X)$.
It has been shown~\cite{qc-bracket} that the dynamical evolution
of a quantum-classical operator $\hat{\chi}(X)$
is given by
\begin{eqnarray}
\partial_t \hat{\chi}(X)&=&\frac{i}{\hbar}[\hat{H},\hat{\chi}(X)]
-\frac{1}{2}\{\hat{H},\hat{\chi}(X)\}+\frac{1}{2}\{\hat{\chi}(X),\hat{H}\}
\nonumber \\
&=&(\hat{H},\hat{\chi}(X)).
\end{eqnarray}
The last equality defines the quantum-classical bracket
in terms of the commutator and the symmetrized
sum of Poisson brackets.

Exploiting what has been done in Refs.~\cite{sergi,sergi2}
for the Poisson bracket and in the previous section
for the commutator,
the quantum-classical bracket can be easily recasted in matrix
form as a non-Hamiltonian commutator.
To this end, one can introduce the operator $\Lambda$
defined in such a way that applying its negative on any pair
of quantum-classical operators
functions $\hat{\chi}_{\alpha}(X)$ and
 $\hat{\chi}_{\nu}(X)$ their Poisson bracket is obtained
\begin{equation}
\{\hat{\chi}_{\alpha},\hat{\chi}_{\nu}\}=
-\hat{\chi}_{\alpha}(X)\Lambda\hat{\chi}_{\nu}(X)=
\sum_{i,j=1}^{2N}\frac{\partial \hat{\chi}_{\alpha}}{\partial X_i}{\cal B}_{i j}
\frac{\partial \hat{\chi}_{\nu}}{\partial X_j}.
\label{Lambda}
\end{equation}

The quantum-classical law of motion can be rewritten as
\begin{eqnarray}
\partial_t \hat{\chi}_{\alpha}
&=&
\frac{i}{\hbar}
\left[\begin{array}{cc} \hat{H} & \hat{\chi}_{\alpha}\end{array}\right]
\cdot\mbox{\boldmath$\cal B$}\cdot
\left[\begin{array}{c} \hat{H} \\ \hat{\chi}_{\alpha} \end{array} \right]
\nonumber\\
&+&
\left[\begin{array}{cc} \hat{H} & \hat{\chi}_{\alpha} \end{array}\right]
\cdot\left[\begin{array}{cc}0 &\frac{\Lambda}{2}\\
    -\frac{\Lambda}{2}& 0\end{array}\right]\cdot
\left[\begin{array}{c}\hat{H}  \\ \hat{\chi}_{\alpha}\end{array}\right].
\nonumber \\
\end{eqnarray}
A more compact form is readily found by defining
 the antisymmetric matrix super-operator
\begin{equation}
\mbox{\boldmath$\cal D$}=\left[\begin{array}{cc} 0& 1+\frac{\hbar\Lambda}{2i}\\
-\left(1+\frac{\hbar\Lambda}{2i}\right) & 0\end{array}\right].
\label{D}
\end{equation}
Using the matrix super-operator in Eq.~(\ref{D})
the quantum-classical law of motion becomes
\begin{equation}
\partial_t \hat{\chi}_{\alpha}=\frac{i}{\hbar}
\left[\begin{array}{cc} \hat{H} & \hat{\chi}_{\alpha} \end{array}\right]
\cdot\mbox{\boldmath$\cal D$}\cdot
\left[\begin{array}{c} \hat{H} \\ \hat{\chi}_{\alpha} \end{array}\right]
=(\hat{H},\hat{\chi}_{\alpha})
=i\hat{\cal L}\hat{\chi}_{\alpha},
\label{qclm}
\end{equation}
where the last equality introduces the quantum-classical Liouville operator
in terms of the quantum-classical bracket.
The structure of Eq.~(\ref{qclm})
is that of the non-Hamiltonian commutator given in
Eq.~(\ref{eq:gen-qlm}) and as such
generalizes the standard quantum laws of motion of Eq~(\ref{qlm}).
It is clear from its definition in Eq.~(\ref{D}) that the antisymmetric
matrix super-operator $\mbox{\boldmath$\cal D$}$
 has not a simple symplectic structure
as $\mbox{\boldmath$\cal B$}$. It contains the operator $\Lambda$ defined in 
Eq.~(\ref{Lambda}) which, in this case, has a symplectic structure.
As such $\mbox{\boldmath$\cal D$}$ introduces a novel mathematical
structure that characterizes the time evolution of quantum-classical
systems.

The Jacobi relation in quantum-classical dynamics is
\begin{equation}
{\cal J}=(\hat{\chi}_{\alpha},(\hat{\chi}_{\nu},\hat{\chi}_{\sigma}) )
+( \hat{\chi}_{\sigma},(\hat{\chi}_{\alpha},\hat{\chi}_{\nu}) )
+( \hat{\chi}_{\nu}, (\hat{\chi}_{\sigma},\hat{\chi}_{\alpha}) ).
\label{qc-jacobi}
\end{equation}
Using the matrix formalism introduced it is simple
to calculate ${\cal J}$ explicitly and to this aim one can consider
the first term on the right hand side of Eq.~(\ref{qc-jacobi}).
The other two terms on the right hand side of Eq.~(\ref{qc-jacobi})
can then be easily calculated
by considering the even permutations 
of $\hat{\chi}_{\alpha}$, $\hat{\chi}_{\nu}$, $\hat{\chi}_{\sigma}$
in the formula obtained for 
$( \hat{\chi}_{\alpha},(\hat{\chi}_{\nu},\hat{\chi}_{\sigma}) )$.
Finally, collecting the terms together one gets
\begin{eqnarray}
{\cal J}&=&
\frac{1}{4}
\{
\hat{\chi}_{\alpha}\Lambda (\hat{\chi}_{\nu}\Lambda \hat{\chi}_{\sigma}) 
-\hat{\chi}_{\alpha}\Lambda (\hat{\chi}_{\sigma}\Lambda \hat{\chi}_{\nu}) 
-(\hat{\chi}_{\nu}\Lambda \hat{\chi}_{\sigma})\Lambda \hat{\chi}_{\alpha} 
\nonumber \\
&+&(\hat{\chi}_{\sigma}\Lambda \hat{\chi}_{\nu})\Lambda \hat{\chi}_{\alpha}
+ \hat{\chi}_{\sigma}\Lambda (\hat{\chi}_{\alpha}\Lambda \hat{\chi}_{\nu}) 
-\hat{\chi}_{\nu}\Lambda(\hat{\chi}_{\alpha}\Lambda \hat{\chi}_{\sigma}) 
\nonumber \\
&-&(\hat{\chi}_{\alpha}\Lambda \hat{\chi}_{\nu})\Lambda \hat{\chi}_{\sigma} 
+ (\hat{\chi}_{\alpha}\Lambda \hat{\chi}_{\sigma})\Lambda \hat{\chi}_{\nu}
+ \hat{\chi}_{\nu}\Lambda (\hat{\chi}_{\sigma}\Lambda \hat{\chi}_{\alpha})
\nonumber \\
&-&\hat{\chi}_{\sigma}\Lambda (\hat{\chi}_{\nu}\Lambda \hat{\chi}_{\alpha}) 
-(\hat{\chi}_{\sigma}\Lambda \hat{\chi}_{\alpha})\Lambda \hat{\chi}_{\nu} 
+(\hat{\chi}_{\nu}\Lambda \hat{\chi}_{\alpha})\Lambda \hat{\chi}_{\sigma}
\}\;.
\nonumber \\
\end{eqnarray}
In order to easily get such expression the following relation 
\begin{eqnarray}
\sum_{i,j=1}^{2N}
\frac{\partial\hat{\chi}_{\alpha}}{\partial X_{i}}
{\cal B}_{i j}
\partial_j (\hat{\chi}_{\nu}\hat{\chi}_{\sigma})&=&
\sum_{i,j=1}^{2N}
\frac{\partial\hat{\chi}_{\alpha}}{\partial X_{i}}
{\cal B}_{i j}
(\partial_j \hat{\chi}_{\nu})\hat{\chi}_{\sigma}
\nonumber \\
&+&
\frac{\partial\hat{\chi}_{\alpha}}{\partial X_{i}}
{\cal B}_{i j}
\hat{\chi}_{\nu}(\partial_j\hat{\chi}_{\sigma})
\end{eqnarray}
was exploited.
Thus it is found that the Jacobi relation does not hold
globally for all points $X$ of phase space
($ {\cal J}\neq 0$).

\section{Nos\'e Dynamics in 
Quantum-Classical Systems}\label{sec:nose}

The antisymmetric matrix $\mbox{\boldmath$\cal B$}$
 enters through $\Lambda$
in the definition
of $\mbox{\boldmath$\cal D$}$.
Following the work of Ref.~\cite{sergi,sergi2} 
thermodynamic constraints can be imposed on the classical bath degrees of 
freedom in quantum-classical dynamics just by modifying the
matrix $\mbox{\boldmath$\cal B$}$.
For clarity it will be explicitly shown how to generalize
the derivation of quantum-classical equations of motion
in the case of  Nos\'e constant 
temperature dynamics~\cite{nose}.
In the Nos\'e case the bath degrees of freedom will be
\begin{equation}
X\equiv (R,\eta,P,p_{\eta})\;,
\end{equation}
$\eta$ and $p_{\eta}$ are the Nos\'e coordinate and momentum.
The following quantum-classical Hamiltonian is assumed
\begin{equation}
\hat{H}_{\rm N}=\hat{K}+\frac{P^2 }{2M}+\frac{p_{\eta}^2 
}{2m_{\eta}}+
\hat{\Phi}(\hat{\chi},R) + gk_BT\eta \;,
\end{equation}
where $\hat{K}$ is the quantum kinetic operator,
$\hat{\Phi}$ is the potential operator coupling classical
and quantum variables, $M$ is the mass of the classical
degrees of freedom, $m_{\eta}$ is Nos\'e inertial parameter,
and $g$ is a numerical constant
whose value (as it will be shown)
must be set equal to the number $ N$
of classical momenta $P$
if one wants to obtain a sampling
of the $R,P$ coordinates in the canonical ensemble.
Then the matrix $\mbox{\boldmath$\cal B$}^{\rm N}$ is
\begin{equation}
\mbox{\boldmath$\cal B$}^{\rm N}=
\left[
\begin{array}{cccc} 0  &  0 & 1 &  0 \\
                    0  &  0 & 0 &  1 \\
		   -1  &  0 & 0 & -P \\
		    0  & -1 & P &  0 
    \end{array}\right]
\;.
\end{equation}
Using $\mbox{\boldmath$\cal B$}^{\rm N}$  the operator $\Lambda_{\rm N}$
and the classical phase space non-Hamiltonian bracket on two generic
variables $A_1$ and $A_2$ can be defined
\begin{equation} 
A_1\Lambda_{\rm N}A_2=-\sum_{i,j=1}^{2N}\frac{\partial A_1}{\partial X_i}
{\cal B}_{i j}^{\rm N}\frac{\partial A_2}{\partial X_j}.
\label{gen-bracket}
\end{equation}
The explicit form of the matrix operator
that defines through Eq.~(\ref{qclm}) the quantum-classical bracket
and the law of motion is then given by
\begin{equation}
\mbox{\boldmath$\cal D$}^{\rm N}=
\left[
\begin{array}{cc} 0 & 1+\frac{\hbar}{2i}\Lambda_{\rm N}\\
-\left(1+\frac{\hbar}{2i}\Lambda_{\rm N}\right) & 0
\end{array}
\right]\;.
\end{equation}

The quantum-classical Nos\'e-Liouville operator is given by
\begin{eqnarray}
\frac{d}{dt}\hat{\chi}&=&i{\cal L}^{\rm 
N}\chi=\frac{i}{\hbar}\left[\begin{array}{cc}\hat{H}_{\rm N} &\chi
 \end{array}\right]
\cdot\mbox{\boldmath$\cal D$}^{\rm N}\cdot
\left[\begin{array}{c} \hat{H}_{\rm N}\\ \hat{\chi} \end{array}\right]
\;.
\label{eqofm}
\end{eqnarray}
One is then led to consider, in the right hand side of (\ref{eqofm}),
the term given by
\begin{eqnarray}
-\hat{H}_{\rm N}\Lambda_{\rm N}\hat{\chi}+\hat{\chi}
\Lambda_{\rm N}\hat{H}_{\rm N}
&=&
\frac{\partial \hat{\Phi}}{\partial R} \frac{\partial \hat{\chi}}{\partial P}
+\frac{\partial \hat{\chi}}{\partial P}\frac{\partial \hat{\Phi}}{\partial R}
-2F_{\eta}\frac{\partial \hat{\chi}}{\partial p_{\eta}}
\nonumber \\
&&
-2\frac{P}{M} \frac{\partial \hat{\chi}}{\partial R}
-2\frac{p_{\eta}}{m_{\eta}} \frac{\partial \hat{\chi}}{\partial \eta}
\nonumber \\
&&
+2\frac{p_{\eta}}{m_{\eta}} P \frac{\partial \hat{\chi}}{\partial P}
\;,
\end{eqnarray}
where $F_{\eta}=\frac{P^2}{M}-gk_BT$.
Finally using the above result
the equation of motions for the dynamical variables are given by
\begin{eqnarray}
\frac{d}{dt}\chi
&=&
\frac{i}{\hbar}\left(H\hat{\chi} -\hat{\chi} H\right)
-\frac{1}{2}\left(\frac{\partial \hat{\chi}}{\partial P}\frac{\partial \Phi}{\partial R}
+\frac{\partial \Phi}{\partial R}\frac{\partial \hat{\chi}}{\partial 
P}\right)
\nonumber \\
&&
+\frac{P}{M}\frac{\partial \hat{\chi}}{\partial R}
+\frac{p_{\eta}}{m_{\eta}}\frac{\partial \hat{\chi}}{\partial \eta}
-\frac{p_{\eta}}{m_{\eta}}P\frac{\partial \hat{\chi}}{\partial P}
+F_{\eta}\frac{\partial \hat{\chi}}{\partial p_{\eta}}
\;.
\nonumber \\
\label{noseeqofm}
\end{eqnarray}

\subsection{Representation in the Adiabatic Basis}

One can express the quantum-classical
equations of motion in the adiabatic states.
Nos\'e quantum-classical Hamiltonian can be rewritten as
\begin{eqnarray}
\hat{H}_{\rm N}&=&\hat{h}(R) + \frac{P^2}{2M}+\frac{p_{\eta}^2}{2m_{\eta}}
+gk_BT\eta \;,
\end{eqnarray}
where it has been introduced the operator 
$\hat{h}(R)=\hat{K}+\hat{\Phi}(\hat{\chi},R)$.
Then the adiabatic states are defined by
\begin{eqnarray}
\hat{h}(R)|\alpha ; R\rangle &=&E_{\alpha}(R)|\alpha ; R\rangle \;.
\end{eqnarray}
In the adiabatic states, Eq.~(\ref{noseeqofm})
 is easily found to be
\begin{eqnarray}
\frac{d}{dt}\chi^{\alpha \alpha '}&=&
i\omega_{\alpha \alpha '}\chi^{\alpha \alpha '}+
\frac{P}{M}\frac{\partial \chi^{\alpha \alpha '}}{\partial R}
\nonumber \\
&&
+ \left(-P\frac{p_{\eta}}{m_{\eta}}\frac{\partial 
}{\partial P}+\frac{p_{\eta}}{m_{\eta}}\frac{\partial}{\partial \eta}
+F_{\eta}\frac{\partial}{\partial p_{\eta}}\right)\chi^{\alpha \alpha '}
\nonumber \\
&&
+\frac{P}{M}d_{\alpha \beta}\chi^{\beta \alpha '}
-\frac{P}{M}\chi^{\alpha \beta '}d_{\beta '\alpha '}
+ \frac{1}{2}\frac{\partial \chi^{\alpha \beta '}}{\partial P}
 F^{\beta ' \alpha '}
 \nonumber \\
 && +\frac{1}{2} F^{\alpha \beta}
 \frac{\partial \chi^{\beta \alpha '}}{\partial P} \;,
 \label{eqofm1}
\end{eqnarray}
where 
$F^{\alpha \beta}=-\langle \alpha |\frac{\partial \Phi}{\partial R}|\beta \rangle$
and $d_{\alpha\beta}=\langle \alpha |\frac{\partial }{\partial R}|\beta \rangle$
is the nonadiabatic coupling vector.
Equation~(\ref{eqofm1}) can be rewritten introducing the Liouville operator
$i{\cal L}^{\rm N}$ such that
\begin{equation}
\frac{d}{dt}\chi^{\alpha \alpha '}=\sum_{\beta \beta `}
i{\cal L}^{\rm N}_{\alpha \alpha ',\beta \beta '}\chi^{\beta \beta '}
\;.
\end{equation}
The operator is 
\begin{eqnarray}
i{\cal L}^{\rm N}_{\alpha \alpha ',\beta \beta '}&=&
i\omega_{\alpha \alpha '}\delta_{\alpha \beta}\delta_{\alpha '\beta '}+
\delta_{\alpha \beta}\delta_{\alpha '\beta '}\frac{P}{M}\frac{\partial }{\partial R}
\nonumber \\
&+&\delta_{\alpha \beta}\delta_{\alpha '\beta '}
\left(-P\frac{p_{\eta}}{m_{\eta}}\frac{\partial 
}{\partial P}+\frac{p_{\eta}}{m_{\eta}}\frac{\partial}{\partial \eta}
+F_{\eta}\frac{\partial}{\partial p_{\eta}}\right)
\nonumber 
\\
&+&\frac{1}{2}\delta_{\alpha \beta} F^{\beta ' \alpha '}\frac{\partial }{\partial P}
 +\frac{1}{2}\delta_{\alpha '\beta '} F^{\alpha \beta}\frac{\partial }{\partial P}
 +\frac{P}{M}d_{\alpha \beta}\delta_{\alpha '\beta '}
\nonumber \\
&-&\frac{P}{M}d_{\beta '\alpha '}\delta_{\alpha \beta}
\;.
\label{liouville1}
\end{eqnarray}
The quantum-classical Liouville operator can be put into
a form that makes its structure more apparent
by adding and subtracting the  term
\begin{equation}
\delta_{\alpha \beta}\delta_{\alpha '\beta '}
\frac{1}{2}\left(F^{\alpha}+F^{\beta '}\right)\frac{\partial}{\partial P}.
\end{equation}
Then using
\begin{equation}
F^{\alpha \beta}=F^{\alpha}+\left(E_{\alpha}-E_{\beta}\right)d_{\alpha \beta}
\end{equation}
and rearranging the terms one obtains
a classical-like Nos\'e-Liouville operator
\begin{eqnarray}
i\hat{L}^{\rm N}_{\alpha \alpha '}&=&
\frac{P}{M}\frac{\partial }{\partial R}
-P\frac{p_{\eta}}{m_{\eta}}\frac{\partial 
}{\partial P}+\frac{p_{\eta}}{m_{\eta}}\frac{\partial}{\partial \eta}
+F_{\eta}\frac{\partial}{\partial p_{\eta}}
\nonumber \\
&+&\frac{1}{2}\left(F^{\alpha}+F^{\alpha '}\right)\frac{\partial}{\partial P}
\end{eqnarray}
and a jump operator
\begin{eqnarray}
-J_{\alpha \alpha ',\beta \beta '}
&=&\delta_{\alpha' \beta'}d_{\alpha \beta}
\left[\frac{P}{M}+\left(\frac{E_{\alpha}-E_{\beta}}{2}\right)\frac{\partial}{\partial 
P}\right]
\nonumber \\
&+&
\delta_{\alpha \beta}d_{\alpha '\beta '}^{*}
\left[\frac{P}{M}+
\left(\frac{E_{\alpha '}-E_{\beta '}}{2}\right)\frac{\partial}{\partial 
P}\right]
\end{eqnarray}
in terms of which the quantum-classical Liouville operator
is finally written as 
\begin{equation}
i{\cal L}^{\rm N}_{\alpha \alpha ',\beta \beta '}
=
i\omega_{\alpha \alpha'}\delta_{\alpha\beta}\delta_{\alpha'\beta'}
+
\delta_{\alpha\beta}\delta_{\alpha'\beta'}i{L}^{\rm N}_{\alpha \alpha '}
-
J_{\alpha \alpha ',\beta \beta '}.
\label{nose-liouville}
\end{equation}
The jump operator $J_{\alpha \alpha ',\beta \beta '}$ is responsible
for transitions between adiabatic states while
the classical-like Nos\'e Liouville operator 
$i{\cal L}^{\rm N}_{\alpha \alpha ',\beta \beta '}$
expresses Nos\'e dynamics on a constant generalized energy surface
with Hellman-Feynman forces given by
$1/2\left(F^{\alpha}+F^{\alpha '}\right)$.

This shows that the matrix form of the non-Hamiltonian commutator 
is suitable for the development of generalized non-Hamiltonian
dynamics for classical degrees of freedom
in quantum-classical systems.

\section{Stationary Nos\'e Density Matrix}\label{sec:stationary}

The average of any operator $\hat{\chi}$
can be calculated from
\begin{equation}
\langle \hat{\chi}\rangle
={\rm Tr}'\int dX~\hat{\rho}_{\rm N}\hat{\chi}(t)
={\rm Tr}'\int dX~\hat{\rho}_{\rm N}
\exp\left(i{\cal L}^{\rm N} t\right)\hat{\chi} \;.
\end{equation}
The action of $\exp\left(i{\cal L}^{\rm N} t\right)$ can be transferred
from $\hat{\chi}$ to $\hat{\rho}_{\rm N}$ by using the cyclic invariance 
of the trace and integrating by parts the terms coming from
the  classical brackets.
One can write
\begin{equation}
i{\cal L}^{\rm N}=\frac{i}{\hbar}\left[\hat{H}_{\rm N},\dots \right]-\frac{1}{2}
\left(\{\hat{H}_{\rm N},\dots\}-\{\dots,\hat{H}_{\rm N}\}\right).
\end{equation}
In this equation the classical bracket terms
are written
\begin{eqnarray}
\{\hat{H}_{\rm N},\dots \}-\{\dots,\hat{H}_{\rm N}\}
&=&
\sum_{i,j=1}^{2N}\left(
\frac{\partial\hat{H}_{\rm N}}{\partial X_i}
{\cal B}_{i j}^N
\frac{\partial \ldots}{\partial X_j}
\right.
\nonumber \\
&-&\left.
\frac{\partial \dots}{\partial X_i}
{\cal B}_{i j}^N
\frac{\partial \hat{H}_{\rm N}}{\partial X_j}\;.
\right)
\end{eqnarray}
When integrating by parts the right hand side,
one obtains a term
proportional to the compressibility
$\kappa_{\rm N}=\sum_{i,j=1}^{2N}\frac{\partial {\cal B}_{i j}^N}{\partial X_i}
\frac{\partial \hat{H}_{\rm N}}{\partial X_j}$.
As a result the mixed quantum-classical Liouville operator, in this 
case,
is not hermitian
\begin{equation}
    \left(i\hat{\cal L}^{\rm N}\right)^{\dag} 
    =-i\hat{\cal L}^{\rm N}-\kappa_{\rm N}\;.
\end{equation}
The average value can then be written as
\begin{equation}
\langle \hat{\chi}\rangle
={\rm Tr}'\int dX~\hat{\chi} \exp\left[-(i{\cal L}^{\rm 
N}+\kappa_{\rm N}) t\right]\hat{\rho}_{\rm N}
\end{equation}
The mixed quantum-classical Nos\'e density matrix
evolves under the equation
\begin{eqnarray}
\frac{\partial}{\partial t}\hat{\rho}_{\rm N}&=&-\frac{i}{\hbar}
\left[\hat{H}^{\rm N},\hat{\rho}_{\rm N}\right]+\frac{1}{2}
\left(\{\hat{H}^{\rm N},\hat{\rho}_{\rm N}\}-\{\hat{\rho}_{\rm N},\hat{H}^{\rm N}\}\right)
\nonumber \\
&-&\kappa_{\rm N}\hat{\rho}_{\rm N}\;.
\label{eq:qc-dme}
\end{eqnarray}

The stationary density matrix $\hat{\rho}_{\rm Ne}$ is defined by
\begin{equation}
(i{\cal L}^{\rm N}+\kappa_{\rm N})\hat{\rho}_{{\rm N}e}=0\;.
\label{eq:qm-case}
\end{equation}
To find the explicit expression one can follow Ref.~\cite{kapral2},
expand the density matrix in powers of $\hbar$
\begin{equation}
\hat{\rho}_{{\rm N}e}=\sum_{n=0}^{\infty}\hbar^n\hat{\rho}_{{\rm N}e}^{(n)}
\;,
\end{equation}
and look for an explicit solution in the adiabatic basis.
In such a basis the Nos\'e-Liouville operator is expressed by
Eq.~(\ref{nose-liouville}) and the Nos\'e Hamiltonian is given by
\begin{eqnarray}
H_{\rm N}^{\alpha}&=&\frac{P^2}{2M}+\frac{p_{\eta}^2}{2m_{\eta}}
+gk_BT\eta+E_{\alpha}(R)\nonumber \\
&=&H^P_{\alpha}(R,P)+\frac{p_{\eta}^2}{2m_{\eta}}+gk_BT\eta
\;.
\end{eqnarray}
Thus one obtains an infinite set of equations corresponding to
the various power of $\hbar$ 
\begin{eqnarray}
iE_{\alpha\alpha'}\rho_{{\rm N}e}^{(0)\alpha\alpha'}&=&0
\label{stat-n=0}
\\
iE_{\alpha\alpha'}\rho_{{\rm N}e}^{(n+1)\alpha\alpha'}&=&
-(iL_{\alpha\alpha'}^{\rm N}+\kappa_{\rm N})
\rho_{{\rm N}e}^{(n)\alpha\alpha'}
\nonumber \\
&+&\sum_{\beta\beta'}J_{\alpha\alpha',\beta\beta'}
\rho_{{\rm N}e}^{(n)\beta\beta'}
~ (n\ge 1)\;.
\label{stat-n+1}
\end{eqnarray}
As shown in Ref.~\cite{kapral2}, in order to ensure that a solution can be found
by recursion, one must discuss the solution of Eq.~(\ref{stat-n+1})
when calculating the diagonal elements $\rho_{{\rm N}e}^{(n)\alpha\alpha}$ in terms
of the off-diagonal ones $\rho_{{\rm N}e}^{(n)\alpha\alpha'}$.
To this end, using
$\rho_{{\rm N}e}^{\prime(n)\alpha\alpha'}=(\rho_{{\rm N}e}^{\prime(n)\alpha'\alpha})^*$,
$J_{\alpha\alpha,\beta\beta'}=J_{\alpha\alpha,\beta'\beta}^*$
and the fact that $J_{\alpha\alpha,\beta\beta}=0$ when
a real basis is chosen,
it is useful to re-write Eq.~(\ref{stat-n+1}) in the form
\begin{equation}
(iL_{\alpha\alpha}^{\rm N}+\kappa_{\rm N})\rho_{{\rm N}e}^{(n)\alpha\alpha}=\sum_{\beta>\beta'}2{\cal R}
\left(J_{\alpha\alpha,\beta\beta'}\rho_{{\rm 
N}e}^{(n)\beta\beta'}\right)\;.
\label{stat-condition}
\end{equation}
One has~\cite{sergi} $(-iL_{\alpha\alpha}^{\rm N}-\kappa_{\rm N})^{\dag}
=iL_{\alpha\alpha}^{\rm N}$. The right hand side of this equation
is expressed by means
of the generalized bracket in Eq.~(\ref{gen-bracket}):
$H_{\rm N}^{\alpha}$ and any general function $f(H_{\rm N}^{\alpha})$
are constants of motion under the action of $iL_{\alpha\alpha}^{\rm N}$.
The phase space compressibility $\kappa_{\rm N}$ associated
with the generalized bracket in the case of Nos\'e dynamics is
\begin{eqnarray}
\kappa_{\rm N}^{\alpha}&=&-\beta\frac{d}{dt}
\left(\frac{P^2}{2M}+\frac{p_{\eta}^2}{2m_{\eta}}
+E_{\alpha}(R)\right)
\nonumber \\
&=&-\beta N\frac{p_{\eta}}{m_{\eta}}=-\beta N\frac{d}{dt}H_T^{\alpha}\;,
\end{eqnarray}
where $ N$ is the number of classical momenta $P$
in the Hamiltonian.
Because of the presence of a non-zero phase space compressibility,
integrals over phase space must be taken using the invariant 
measure~\cite{tuckerman}
\begin{equation}
d{\cal M}=\exp(-w_{\rm N}^{\alpha})dRdPd\eta dp_{\eta}\;,
\end{equation}
where $w_{\rm N}^{\alpha}=\int dt\kappa_{\rm N}^{\alpha}$
is the indefinite integral of the compressibility.
To insure that a solution to Eq.~(\ref{stat-condition})
exists one must invoke the theorem
of Fredholm alternative, requiring that the 
right-hand side of Eq.~(\ref{stat-condition}) 
be orthogonal to the null space of 
$(iL_{\alpha\alpha}^{\rm N})^{\dagger}$~\cite{hilbert}.
The null-space of this operator
consists of functions of the form~\cite{sergi2}
$f(H_{\rm N}^{\alpha})$,
where $f(H_{\rm N}^{\alpha})$ can be any function
of the adiabatic Hamiltonian $H_{\rm N}^{\alpha}$.
Thus the condition to be satisfied is
\begin{equation}
\int d{\cal M}\sum_{\beta>\beta'}2{\cal R}
\left(J_{\alpha\alpha,\beta\beta'}\rho_{{\rm N}e}^{(n)\beta\beta'}\right)
f(H_{\rm N}^{\alpha})=0 \;.
\label{fredholm}
\end{equation}
Apart from the integration on the additional Nos\'e phase space variable
there is no major difference with the proof given in Ref.~\cite{kapral2}:
$2{\cal R}\left(J_{\alpha\alpha,\beta\beta'}\rho_{Ne}^{(n)\beta\beta'}\right)$
and $f(H_{\rm N}^{\alpha})$ are respectively 
an odd and an even function of $P$; 
this guarantees the validity of Eq.~(\ref{fredholm}).

Thus one can write the formal solution of Eq.~(\ref{stat-condition})
as
\begin{equation}
\rho_{{\rm N}e}^{(n)\alpha\alpha}
=(iL_{\alpha\alpha}^{\rm N}+\kappa_{\rm N})^{-1}
\sum_{\beta>\beta'}2{\cal R}
\left(J_{\alpha\alpha,\beta\beta'}\rho_{{\rm 
N}e}^{(n)\beta\beta'}\right)\;,
\label{eq:sol1}
\end{equation}
and the formal solution of Eq.~(\ref{stat-n+1}) for 
$\alpha\neq\alpha'$ as
\begin{eqnarray}
\rho_{{\rm N}e}^{(n+1)\alpha\alpha'}&=&
\frac{i}{E_{\alpha\alpha'}}
(iL_{\alpha\alpha'}^{\rm N}+\kappa_{\rm N})
\rho_{{\rm N}e}^{(n)\alpha\alpha'}
\nonumber \\
&&
-\frac{i}{E_{\alpha\alpha'}}
\sum_{\beta\beta'}J_{\alpha\alpha',\beta\beta'}\rho_{{\rm N}e}^{(n)\beta\beta'}
\;.
\label{eq:sol2}
\end{eqnarray}
Equations~(\ref{eq:sol1}) and~(\ref{eq:sol2}) allows one to calculate
$\rho_{\rm Ne}^{\alpha\alpha'}$ to all orders in $\hbar$
once $\rho_{\rm Ne}^{(0)\alpha\alpha'}$ is given.
This order zero term is obtained by the solution of
$(iL_{\alpha\alpha}^{\rm N}+\kappa_{\rm N})
\rho_{{\rm N}e}^{(0)\alpha\alpha}=0$. All higher order terms
are obtained by the action of $E_{\alpha\alpha'}$, the imaginary unit 
$i$ and $J_{\alpha\alpha'\beta\beta'}$ (involving factors of 
$d_{\alpha\alpha'}$, $P$ and derivatives with respect to $P$.
Hence, one can conclude that functional dependence of
$\rho_{\rm Ne}^{(0)\alpha\alpha}$ on
the Nos\'e variables $\eta$ and $p_{\eta}$
is preserved in higher order terms $\rho_{\rm Ne}^{(n)\alpha\alpha'}$.

One can find a stationary solution
to order $\hbar$
by considering the first two equations of the set
given by Eqs.~(\ref{stat-n=0}) and (\ref{stat-n+1}):
\begin{eqnarray}
\left[\hat{H}_{\rm N},\hat{\rho}_{{\rm N}e}^{(0)}\right]& =&0  
\qquad \qquad  (n=0)\;,
\label{n=0}
\\
i\left[\hat{H}_{\rm N},\hat{\rho}_{{\rm N}e}^{(1)}\right]&=&
-\frac{1}{2}\left(\hat{H}_{\rm N}\Lambda_{\rm N}\hat{\rho}_{{\rm N}e}^{(0)}
-
\hat{\rho}_{{\rm N}e}^{(0)}\Lambda_{\rm N}\hat{H}^{\rm N}\right)
\nonumber \\
&&  (n=1)\;.
\label{n=1}
\end{eqnarray}

For the ${\cal O}(\hbar^0)$ term one can make the ansatz
\begin{equation}
\hat{\rho}_{Ne}^{(0)\alpha\beta}=\frac{1}{Z}e^{w_{\rm N}^{\alpha}}
\delta\left({\cal C}-H^{\alpha}_{\rm N}\right)\delta_{\alpha\beta}
\;, \label{ansatz}
\end{equation}
where $Z$ is
\begin{eqnarray}
Z&=&\sum_{\alpha}\int d{\cal M}~
\delta\left({\cal C}-H^{\alpha}_{\rm N}\right)
\end{eqnarray}
and obtain
\begin{eqnarray}
\hat{\rho}_{Ne}^{(1)\alpha \beta}
&=&-i
\frac{P}{M}d_{\alpha \beta}\hat{\rho}_{Ne}^{(0) \beta}
\left[\frac{1-e^{-\beta(E_{\alpha}-E_{\beta})}}{E_{\beta}-E_{\alpha}}
+\frac{\beta}{2}
\right.
\nonumber \\
&& \left.
\left(1+e^{-\beta(E_{\alpha}-E_{\beta})}\right)
 \right]
 \label{ansatz2}
\end{eqnarray}
for the ${\cal O}(\hbar)$ term.

Equations~(\ref{ansatz}) and (\ref{ansatz2}) give the explicit form
of the stationary solution of the Nos\'e-Liouville equation
up to order ${\cal O}(\hbar)$. 
One can now prove that,
 when calculating averages of quantum-classical operators
depending only on physical phase space variables, ${\cal G}_{\alpha}(R,P)$,
the canonical form of the stationary density is obtained.
It can be noted that it will suffice to prove this result
for the ${\cal O}(\hbar^{^{0}})$ term since,
as discussed before, the differences with
the standard case are  contained  therein.

Indeed, when calculating
\begin{eqnarray}
\langle{\cal G}_{\alpha}(R,P)\rangle
&\propto&=\sum_{\alpha}\int d{\cal M}~{\cal G}_{\alpha}(R,P)
\nonumber \\
&&\times
\delta({\cal C}-H^{\alpha}_{\rm T}-gk_{B}T\eta)\;.
\end{eqnarray}
Considering the delta function integral over Nos\'e variables, one has
\begin{eqnarray}
\int dp_{\eta}d\eta~e^{-N\eta}
\delta({\cal C}&-&H_{\alpha}^T-gk_{B}T\eta)
\nonumber \\
&=&{\rm const}\times
\exp[-\beta( N/g) H_{\alpha}^T(R,P)]\;,
\nonumber \\
\end{eqnarray}
where it has been used the property
$\delta(f(s))=[df/ds]^{-1}_{s=s_0}\delta(s-s_0)$
($s_0$ is the zero of $f(s)$).
Thus, at variance with what found in Ref.~\cite{tosatti},
in order to recover the canonical distribution
in the quantum-classical case, one must set $g= N$
as it is done in the classical case~\cite{nose,hoover}.
If the dynamics is ergodic and if one could integrate
quantum-classical equations of motion
for sufficiently long time, the phase space
integral could be substituted by a time integral
along the trajectory~\cite{ray}.
Ergodicity could be enforced by modern
advanced sampling techniques~\cite{massive-nhc}
but long time stable integration
of quantum-classical dynamics is still
a challenge.

\section{Conclusions and Perspectives}

In this paper 
a generalized non-Hamiltonian form of quantum mechanics
has been presented.
This has been achieved through the introduction
of a suitable non-Hamiltonian commutator
which has been obtained by generalizing the
symplectic structure of the standard
quantum mechanical commutator.
Therefore, it has been demonstrated that a single idea
(i.e. generalizing the symplectic structure
of the bracket while retaining its antisymmetric form)
is able to describe in a unified way non-Hamiltonian theories
both in classical and quantum mechanics.
The non-Hamiltonian form of quantum mechanics
here presented provides a general mathematical
structure which encompasses the ideas proposed
by Weinberg to introduce non-linear effects in quantum mechanics
and whose physical content remains yet to be unveiled.

For the sake of illustrating the possible use of 
non-Hamiltonian commutators, it has been shown that
they subsume the quantum-classical bracket proposed
by other authors. Moreover, their matrix structure
has been used to define Nos\'e  dynamics on the
classical degrees of freedom in quantum-classical systems.
It has been also shown that the non-Hamiltonian quantum-classical bracket
can be easily generalized to treat  other
thermodynamic constraints such as those provided by
barostats or Nos\'e-Hoover chains.
The respective stationary density matrices are easily derived.
The implementation of thermodynamic constraints
for the classical degrees of freedom in quantum-classical systems
could be considered both as a practical and a conceptual improvement.
For example, thermostated dynamics can be useful for preparing
systems into desired initial conditions
or for ensuring a good thermalization of the classical bath degrees of freedom
providing a way to control the nonadiabatic character of the dynamics.
On the conceptual side one could note that, historically,
deterministic dynamics with thermodynamic constraints
for purely classical systems has provided well defined
algorithms to treat open systems
both in and out of equilibrium.
Thus, the possibility to use the same tool in the case
of quantum-classical systems could disclose novel routes to the
numerical study of open quantum systems.

In conclusion, the non-Hamiltonian quantum formalism
introduced in this paper sets a unified framework  
with the non-Hamiltonian classical algebra
and, at the same time, discloses various routes for investigating 
generalized quantum and quantum-classical systems.
Such studies will be performed in the future.

\vspace{1cm}
\begin{flushleft}
{\bf Acknowledgments}
\end{flushleft}
The author is grateful to Raymond
Kapral for suggestions and criticisms.

\appendix

\section{NHC and NPT}\label{sec:nhc-npt}

The calculations of the previous sections show that
the introduction of extended system dynamics on the classical
part of the system amounts
to modify the operator of Eq.~(\ref{D})
by simply substituting the classical bracket
operator given in Eq.~(\ref{Lambda})
with the one suited to express the desired
extended system dynamics~\cite{sergi,sergi2}.

Thus, in order to couple a Nos\'e-Hoover chain~\cite{nhc}
to the classical coordinates,
the classical phase space point is defined as
\begin{equation}
X=(R,\eta_{1},\eta_{2},P,p_{\eta_1},p_{\eta_2})
\;,
\end{equation}
where for simplicity one is considering a chain of just two
thermostat coordinates $\eta_{1}$, $\eta_{2}$ and momenta
$p_{\eta_1}$, $p_{\eta_2}$.
\begin{eqnarray}
\hat{H}^{\rm NHC}&=&\frac{\hat{p}^2 }{2m}+\frac{P^2 }{2M}
+\frac{p_{\eta_1}^2 }{2m_{\eta_1}}
+\frac{p_{\eta_2}^2 }{2m_{\eta_2}}
\nonumber \\
&+&\hat{\Phi}(\hat{q},R) + gk_BT\eta_1 +gk_BT\eta_2
\;,
\end{eqnarray}
where $m_{\eta_1}$ and $m_{\eta_2}$ are the inertial
parameters of the thermostat variables.
As shown in  Ref.~\cite{sergi,sergi2},
one can define an antisymmetric matrix
\begin{equation}
\mbox{\boldmath$\cal B$}^{\rm NHC}=
\left[
\begin{array}{cccccc}
0 & 0 & 0 & 1 & 0 & 0 \\
0 & 0 & 0 & 0 & 1 & 0 \\
0 & 0 & 0 & 0 & 0 & 1 \\
-1 & 0 & 0 & 0 & -P & 0 \\
0 & -1 & 0 & P& 0 & -p_{\eta_1}\\
0 & 0 & -1 & 0 & p_{\eta_1} & 0
\end{array}
\right]\;.
\end{equation}
The matrix $\mbox{\boldmath$\cal B$}^{\rm NHC}$ 
determines
the operator $\Lambda^{\rm NHC}$ 
which in turn 
provides the
the non-Hamiltonian bracket
according to Eq.~(\ref{Lambda}).
The Nos\'e-Hoover chain classical equations of motion
in phase space~\cite{sergi} are then given by
\begin{equation}
\dot{X}=-X\Lambda^{\rm NHC}\hat{H}^{\rm NHC}.
\end{equation}
Quantum-classical dynamics is then introduced
using the matrix super-operator
\begin{equation}
\mbox{\boldmath$\cal D$}^{\rm NHC}=\left[\begin{array}{cc}
0 & 1+\frac{\hbar}{2i}\Lambda^{\rm NHC}\\
-\left(1+\frac{\hbar}{2i}\Lambda^{\rm NHC}\right) & 0\end{array}\right].
\end{equation}
As previously shown by means of the latter
the quantum-classical equations of motion
are then given by
\begin{eqnarray}
\frac{d\hat{\chi}}{dt}&=&\frac{i}{\hbar}
\left[\begin{array}{cc}\hat{H}^{\rm NHC} &\hat{\chi}\end{array}\right]
\cdot\mbox{\boldmath$\cal D$}^{\rm NHC}\cdot
\left[\begin{array}{c}\hat{H}^{\rm NHC} \\ \hat{\chi}\end{array}\right].
\end{eqnarray}
The equations of motion can be represented using the
adiabatic basis obtaining the Liouville super-operator
\begin{eqnarray}
i{\cal L}_{\alpha\alpha',\beta\beta'}^{\rm NHC}&=&
(i\omega_{\alpha\alpha'}+iL_{\alpha\alpha'}^{\rm NHC})\delta_{\alpha\beta}
\delta_{\alpha'\beta'} -J_{\alpha\alpha',\beta\beta'}
\;, \nonumber \\
\end{eqnarray}
where
\begin{eqnarray}
iL_{\alpha\alpha'}^{\rm NHC}&=&\frac{P}{M}\frac{\partial}{\partial R}
+\frac{1}{2}(F^{\alpha}+F^{\alpha'})\frac{\partial}{\partial P}
\nonumber \\
&+&\sum_{k=1}^2(\frac{p_{\eta_k}}{m_{\eta_k}}\frac{\partial}{\partial\eta_k}
+F_{\eta_k}\frac{\partial}{\partial p_{\eta_k}})
\nonumber \\
&-& \frac{p_{\eta_2}}{m_{\eta_2}}
p_{\eta_1}\frac{\partial}{\partial p_{\eta_1}}
\;,
\end{eqnarray}
with $F_{\eta_2}=p_{\eta_1}^2/m_{\eta_1}-gk_BT$.
The proof of the existence of stationary density matrix 
in the case of Nos\'e-Hoover chains follows
the same logic of the simple Nos\'e-Hoover case.
In the adiabatic basis
the density matrix stationary up to order bar
has the same form as given in Eqs.~(\ref{ansatz})
and~(\ref{ansatz2}). One has just to replace
Eq.~(\ref{ansatz}) for the order zero term
with
\begin{eqnarray}
\rho_{{\rm NHC}e}^{(0)\alpha\beta}&=&\frac{1}{Z}
e^{-\beta\left[\frac{P^2}{2M}+E_{\alpha}(R)
+\sum_{k=1}^2\left(\frac{p_{\eta_k}^2}{2m_{\eta_k}}
+gk_BT\eta_k\right)\right]}
\nonumber \\
\end{eqnarray}
with obvious definition of $Z$.

For the case of constant pressure and temperature dynamics,
the equations of motion treated
in Ref~\cite{ribes} are here considered. This time,
the extended phase space point is 
\begin{equation}
X=(R,\eta,V,P,p_{\eta},p_V)
\;,
\label{xnpt}
\end{equation}
and the Hamiltonian quantum-classical operator is
\begin{eqnarray}
\hat{H}^{\rm NPT}&=&\frac{\hat{p}^2 }{2m}+\frac{P^2 }{2M}
+\frac{p_{\eta}^2 }{2m_{\eta_1}}
+\frac{p_V^2 }{2m_V}
\nonumber \\
&+&\hat{\Phi}(\hat{q},R) + gk_BT\eta +P_{\rm ext}V
\;.
\end{eqnarray}
The equations of motion for the classical coordinates
are~\cite{sergi}
\begin{eqnarray}
\dot{R}&=&\frac{P}{m_i}+R\frac{p_V}{3Vm_V}
\label{npt-first}\\
\dot{p}_{\eta}&=&\frac{p_{\eta}}{m_{\eta}}\\
\dot{V}&=&\frac{p_V}{m_V}\\
\dot{P}&=&-\frac{\partial \Phi}{\partial R}
-P\frac{p_V}{3Vm_V}
-P\frac{p_{\eta}}{m_{\eta}}\\
\dot{p}_{\eta}&=&
\sum_{i=1}^N\frac{P^2}{m_i}+\frac{p_V^2}{m_V}-gk_BT\\
\dot{p}_V&=&F_V -p_V\frac{p_{\eta}}{m_{\eta}}
\label{npt-last}
\end{eqnarray}
with
\begin{eqnarray}
F_V&=&\frac{1}{3V}\left[\sum_{i=1}^N\frac{P^2}{M}
-\frac{\partial \Phi}{\partial R}\cdot R\right]-P_{ext}
\;.
\end{eqnarray}
The antisymmetric matrix to define the
Operator $\Lambda^{\rm NPT}$ in Eq.~(\ref{Lambda})
is then
\begin{equation}
\mbox{\boldmath$\cal B$}^{\rm NPT}=
\left[
\begin{array}{cccccc}
0 & 0 & 0 & 1 & 0 & \frac{R}{3V}\\
0 & 0 & 0 & 0 & 1 & 0 \\
0 & 0 & 0 & 0 & 0 & 1 \\
-1 & 0 & 0 & 0 & -P & -\frac{P}{3V}\\
0 & -1 & 0 & P & 0 & p_V \\
-\frac{R}{3V} & 0 & -1 & \frac{P}{3V} & -p_V & 0
\end{array}
\right]
\;.
\label{bnpt}
\end{equation}
The matrix $\mbox{\boldmath$\cal B$}^{\rm NPT}$ is the same
as that given in Ref.~\cite{sergi} but this time the order
of coordinates and momenta in the classical
extended phase space point definition of Eq.~(\ref{xnpt})
ensures that $\Lambda^{\rm NPT}$ in Eq.~(\ref{Lambda})
makes the equations of motion
\begin{equation}
\dot{X}=-X\Lambda^{\rm NPT}\hat{H}^{\rm NPT}
\end{equation}
exactly equivalent to Eqs.~(\ref{npt-first}-\ref{npt-last}).
Then one can define the matrix operator 
$\mbox{\boldmath$\cal D$}^{\rm NPT}$ and write down
the quantum-classical equations of motion.
In the adiabatic basis the equations are written by means of the
Liouville operator
\begin{eqnarray}
i{\cal L}_{\alpha\alpha',\beta\beta'}^{\rm NPT}&=&
(i\omega_{\alpha\alpha'}+iL_{\alpha\alpha'}^{\rm NPT})\delta_{\alpha\beta}
\delta_{\alpha'\beta'} -J_{\alpha\alpha',\beta\beta'}
\;,
\nonumber \\
\end{eqnarray}
with
\begin{eqnarray}
iL_{\alpha\alpha'}^{\rm NPT}&=&iL_{\alpha\alpha'}^{\rm NH}
+\frac{P_V}{3Vm_V}R\frac{\partial}{\partial R}
-\frac{P_V}{3Vm_V}P\frac{\partial}{\partial P}
\nonumber \\
&+&\frac{P_V}{m_V}\frac{\partial}{\partial V}
+(F_V-\frac{p_{\eta}}{m_{\eta}})\frac{\partial}{\partial P_V}
\;.
\end{eqnarray}
The stationary density matrix is derived as usual
and in the adiabatic basis is expressed again in the form given
by Eqs.~(\ref{ansatz}) and~(\ref{ansatz2})
with the ${\cal O}(\hbar^0)$ term given by
\begin{eqnarray}
\rho_{{\rm NPT}e}^{(0)\alpha\beta}&=&\frac{1}{Z}
e^{-\beta\left[\frac{P^2}{2M}+E_{\alpha}(R)
+\frac{p_{\eta}^2}{2m_{\eta}}
+gk_BT\eta +\frac{P_V^2}{2m_V}+P_{ext}V\right]}
\;.
\nonumber \\
\end{eqnarray}


\end{document}